%% file: RJwrapper.tex
\documentclass[a4paper]{report}
\usepackage[utf8]{inputenc}
\usepackage[T1]{fontenc}
\usepackage{RJournal}
\usepackage{amsmath,amssymb,array}
\usepackage{booktabs}

\providecommand{\tightlist}{%
  \setlength{\itemsep}{0pt}\setlength{\parskip}{0pt}}

\usepackage{longtable}

\makeatletter
 \let\@cite@ofmt\@firstofone
 \def\@biblabel#1{}
 \def\@cite#1#2{{#1\if@tempswa , #2\fi}}
\makeatother
\newlength{\cslhangindent}
\newlength{\csllabelwidth}
 {\begin{list}{}{%
  \setlength{\itemindent}{0pt}
  \setlength{\leftmargin}{0pt}
  \setlength{\parsep}{0pt}
  \ifodd #1
   \setlength{\leftmargin}{\cslhangindent}
   \setlength{\itemindent}{-1\cslhangindent}
  \fi
  \setlength{\itemsep}{#2\baselineskip}}}
 {\end{list}}
\usepackage{calc}

\begin{document}

\sectionhead{Contributed research article}
\volume{XX}
\volnumber{YY}
\year{20ZZ}
\month{AAAA}

\begin{article}
  \input{moveEZ}
\end{article}

\end{document}

%% file: moveEZ.tex
\title{moveEZ: An R Package for Animated Biplots}

\author{by Raeesa Ganey and Johané Nienkemper-Swanepoel}

\maketitle

\abstract{%
The moveEZ (pronounced move easy) R package provides tools for constructing animated Principal Component analysis (PCA) biplots that represent samples and variables simultaneously. These visualizations reveal how multivariate structures evolve across the ordered levels of a categorical variable. Built as an extension to the biplotEZ package, moveEZ offers three animation frameworks of increasing methodological complexity: a fixed variable frame, in which variable vectors remain constant and only sample positions are animated; and two dynamic frames, in which both sample positions and variable vectors are recomputed and animated at each level. The dynamic frames support Generalized Procrustes analysis (GPA) for alignment and reflection to ensure visual continuity across levels, and are compatible with high-dimensional datasets including grouped structures. The package integrates with gganimate to produce high-quality animations suitable for publications and presentations, and supports both animated and static faceted displays via a single argument. Although originally motivated by tracking shifts in African climate indicators, moveEZ is domain-agnostic and applicable wherever multivariate measurements are recorded repeatedly across an ordered categorical variable, including economic, ecological, and biological settings.
}

\section{Introduction}\label{introduction}

The \CRANpkg{moveEZ} R package introduces animated Principal Component analysis (PCA) biplots for the visualization of multivariate data observed repeatedly across the ordered levels of a categorical variable. PCA biplots, constructed via the companion package \CRANpkg{biplotEZ} (\citet{biplotEZ}), are effective for revealing structure in high-dimensional data. However, when data are sequentially structured in this way, restricting visualizations to individual levels obscures the continuity of change, making gradual movement in samples or variables across the sequence difficult to detect. \CRANpkg{moveEZ} addresses this by animating transitions between levels and supporting dynamic exploration, allowing analysts to perceive movement in multivariate space as a continuous process rather than a sequence of discrete comparisons. Although originally motivated by the tracking of climate conditions over time, the package is domain-agnostic and applicable wherever multivariate measurements are recorded repeatedly across an ordered categorical variable; whether that variable represents time, experimental stages, algorithmic iterations, or any other meaningful sequence.

\section{PCA Biplots}\label{pca-biplots}

The aim is to approximate a high-dimensional multivariate dataset in two dimensions for visualization. Consider a dataset \({\bf{X}}\) comprising \(n\) observations and \(p\) continuous variables, along with an additional categorical variable referred to throughout this paper as the time variable whose ordered levels define the sequential structure in the data. This variable does not need to correspond to chronological time; it may represent any ordered index such as experimental stages, algorithmic iterations, or measurement occasions. Each distinct level of the time variable partitions \({\bf{X}}\) into a time slice: the subset of rows corresponding to that level. The collection of time slices across all levels constitutes the full sequential dataset that \CRANpkg{moveEZ} animates.

Where appropriate, standardization (centering and scaling) is applied prior to dimension reduction. Singular value decomposition (SVD) of \({\bf{X}}\) yields:
\[{\bf{X=UDV'}}.\]
A standard PCA biplot represents samples in principal coordinates, taken from the first two columns of \({\bf{UD}}\), and variables in standard coordinates, taken from the first two columns of \({\bf{V}}\) (\citet{Greenacre2022PCA}).

Variables in a PCA biplot can be represented either as directed vectors, as introduced by (\citet{Gabriel}), or as calibrated axes, which support interpolation and prediction of sample values (\citet{Blasius}, \citet{Biplots}, \citet{UB}). Since \CRANpkg{moveEZ} produces dynamic visualizations across multiple time points, directed vectors are preferred over calibrated axes. They produce a less cluttered display, which eases visual tracking of subtle changes in both samples and variables as the animation progresses.

\section{\texorpdfstring{The \CRANpkg{moveEZ} framework}{The  framework}}\label{the-framework}

\CRANpkg{moveEZ} animates PCA biplots across the levels of an ordered variable, the time variable, by interpolating between successive biplot states using a linear easing function, \texttt{ease\_aes()} from the \CRANpkg{tweenr} package (\citet{tweenr}), applied via the \texttt{transition\_states()} function in the package \CRANpkg{gganimate} (\citet{gganimate}). Linear easing produces transitions at a constant rate, ensuring smooth and visually interpretable movement between states. A brief pause at each state allows the user to clearly identify the biplot corresponding to each time level before the animation continues.

The package provides three functions: \texttt{moveplot()}, \texttt{moveplot2()}, and \texttt{moveplot3()}, each implementing a different conceptual approach to animating the biplot across time. The key distinction between them concerns whether the PCA solution is computed once on the full dataset or separately for each time slice, and consequently whether the variable vectors remain fixed or evolve over time. All three functions share a \texttt{move} argument that toggles between a dynamic animation (\texttt{move\ =\ TRUE}) and a static faceted display (\texttt{move\ =\ FALSE}), the latter being useful for figures in publications or detailed comparison across time points. Table
\ref{tab:comparison-table} summarizes the key differences between the three functions to assist users in selecting the most appropriate approach for their data and research question.

\begin{table}[!h]
\centering
\caption{\label{tab:comparison-table}Comparison of moveplot(), moveplot2(), and moveplot3().}
\centering
\resizebox{\ifdim\width>\linewidth\linewidth\else\width\fi}{!}{
\fontsize{9}{11}\selectfont
\begin{tabular}[t]{>{\raggedright\arraybackslash}p{3cm}>{\raggedright\arraybackslash}p{3.5cm}>{\raggedright\arraybackslash}p{3.5cm}>{\raggedright\arraybackslash}p{3.5cm}}
\toprule
Feature & moveplot() & moveplot2() & moveplot3()\\
\midrule
PCA computed & Once, on full dataset & Per time slice & Per time slice\\
Variable vectors & Fixed & Dynamic & Dynamic\\
Sample coordinates & Dynamic & Dynamic & Dynamic\\
Alignment & Not required & Manual (align.time, reflect) & Automated (GPA)\\
Min. observations per time slice & One & Multiple & Multiple\\
\addlinespace
Balance across time slices & Not required & Not required & Required\\
Best suited for & Stable variance structure; single observations per group per time & Evolving variance structure; manual alignment feasible & Evolving variance structure; many time slices\\
\bottomrule
\end{tabular}}
\end{table}

Table \ref{tab:comparison-table} shows that there is a trade-off between simplicity and advanced methodological choices which is guided by the data. \texttt{moveplot()} assumes that the principal components computed on the full dataset provide a meaningful basis for projecting each time slice. This is appropriate when the underlying variance--covariance structure is stable across time. When this assumption does not hold, that is, when the directional relationships among variables themselves evolve then \texttt{moveplot2()} or \texttt{moveplot3()} are more appropriate, as they compute a separate PCA solution for each time slice. The distinction between these two is one of practicality: \texttt{moveplot2()} requires the user to manually identify and correct orientation discontinuities between frames, while \texttt{moveplot3()} automates this process using Generalized Procrustes analysis (GPA), making it the preferred choice when the number of time levels is large or when discontinuities are difficult to detect visually.

In all three functions, the core biplot construction is handled by \CRANpkg{biplotEZ} (\citet{biplotEZ}), and the resulting object is passed directly to the relevant moveplot function via the pipe operator.

\section{Fixed variable frame}\label{fixed-variable-frame}

\begin{figure}[!]
\includegraphics[width=1\linewidth]{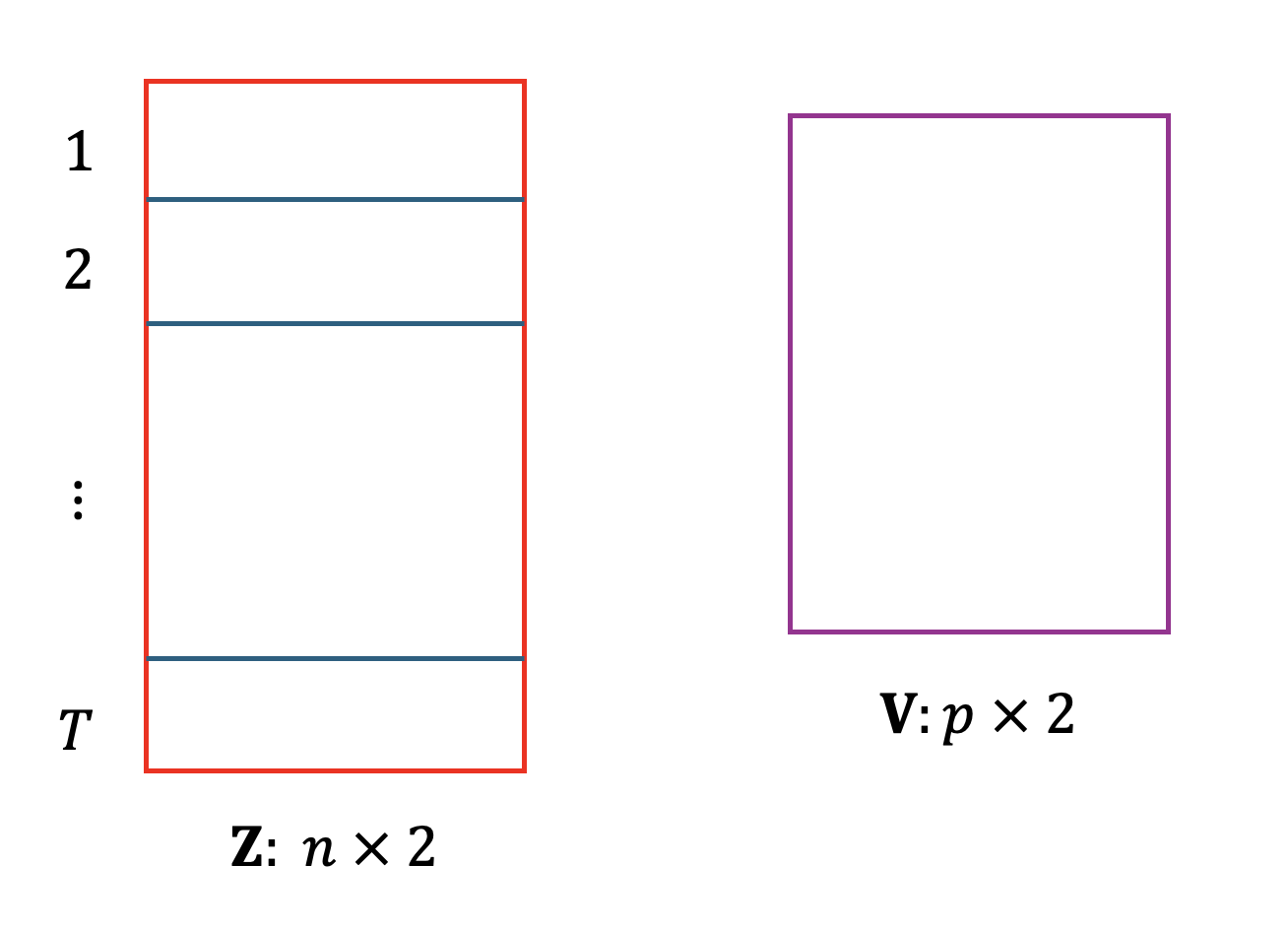} \caption{The fixed variable framework where the sample coordinates in matrix ${\bf{Z}}$ are sliced according to the levels of the time variable $T$ and the variable vectors in matrix ${\bf{V}}$ remaining fixed.}\label{fig:fixed-frame}
\end{figure}

In the fixed variable frame, a single PCA solution is performed on the full dataset \({\bf{X}}\), yielding one set of sample scores, \({\bf{Z}}\), represented by plotted coordinates, and one set of variable loadings, \({\bf{V}}\), as vectors. The variable vectors remain constant throughout the animation, providing a stable reference frame for interpretation. The sample scores are then sliced according to the levels of the time variable and animated sequentially, highlighting how observations move relative to a fixed coordinate system defined by the overall variance structure of the data. This framework is illustrated in Figure \ref{fig:fixed-frame}.

This approach is most appropriate when the underlying variance--covariance structure can be assumed stable across time, and is the only viable option when there is a single observation per group per time level, since a per-slice PCA solution would not be feasible in that setting.

\subsection{\texorpdfstring{Demonstration: \emph{gapminder} data}{Demonstration: gapminder data}}\label{gapminderdata}

To illustrate \texttt{moveplot()} with a single observation per group per time level, we use the \emph{gapminder} dataset available in the \CRANpkg{gapminder} package (\citet{gapminder}), which contains continent-level statistics on life expectancy (lifeExp), population size (pop), and GDP per capita measured every five years from 1952 to 2007 (gdpPercap). For simplicity, the data are aggregated by continent and year, yielding one observation per continent per year:

\begin{verbatim}
data("gapminder")

gapminder <- gapminder |> mutate(year = as.factor(year))

avg_year_continent <- gapminder |>
  group_by(year, continent) |>
  summarise(
    across(c(lifeExp, pop, gdpPercap), \(x) round(mean(x, na.rm = TRUE), 2)),
    .groups = "drop")
\end{verbatim}

A PCA biplot is first constructed using \CRANpkg{biplotEZ}, with continent as the grouping aesthetic and year as the sample label. Because the three continuous variables are on vastly different scales, standardization is applied:

\begin{verbatim}
bp <- biplot(avg_year_continent, scaled = TRUE) |>
  PCA(group.aes = avg_year_continent$continent) |>
  samples(label.name = avg_year_continent$year) |>
  legend.type(samples = TRUE) |>
  plot()
\end{verbatim}

\begin{figure}[!]
\includegraphics[width=1\linewidth]{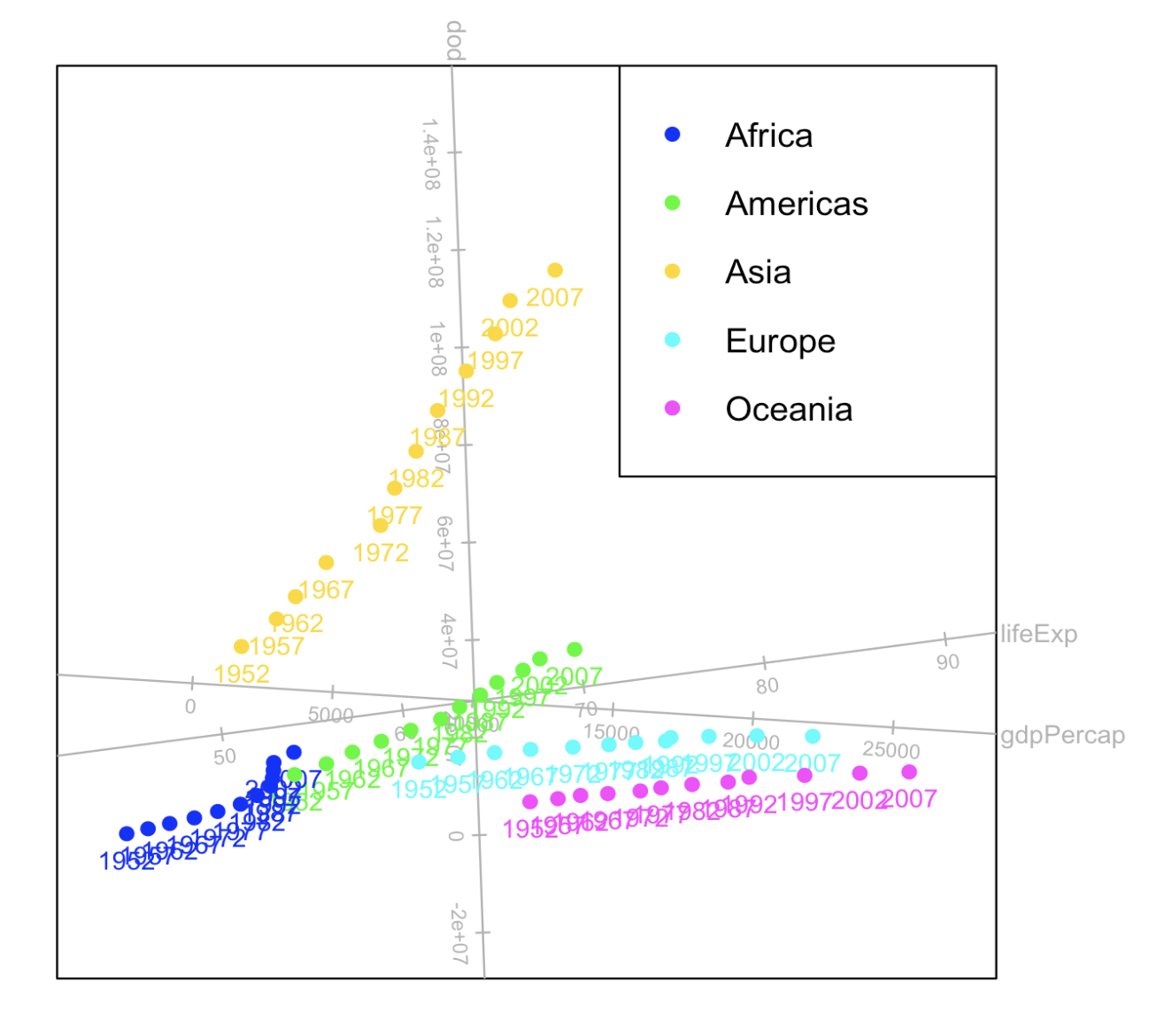} \caption{PCA biplot of the gapminder data colored by continent and labeled by year.}\label{fig:biplot-gap}
\end{figure}

The resulting static biplot, shown in Figure \ref{fig:biplot-gap}, reflects the overall variance structure across all time points. This biplot object is then passed to \texttt{moveplot()}:

\begin{verbatim}
bp |> moveplot(
    time.var = "year",
    group.var = "continent",
    hulls = FALSE,
    scale.var = 3,
    move = FALSE)
\end{verbatim}

The key arguments controlling the visual output of \texttt{moveplot()} are as follows. The \texttt{hulls} argument governs whether groups are represented by convex hulls (\texttt{hulls\ =\ TRUE}, the default) or individual sample points (\texttt{hulls\ =\ FALSE}). Convex hulls are preferable when each group contains many observations and the interest lies in tracking group-level movement, as they provide a compact summary of regional spread at each time point. Individual points are more appropriate when sample-level trajectories are of interest, or when group sizes are small. The \texttt{shadow} argument, available only when \texttt{hulls\ =\ FALSE}, retains faded traces of previous sample positions in the animation, leaving visible trails that convey the direction and relative speed of movement across time. The \texttt{scale.var} argument applies a numeric multiplier to the variable vectors, which can improve interpretability when default vector lengths are too short to convey meaningful directionality.

\begin{figure}[!]
\includegraphics[width=1\linewidth]{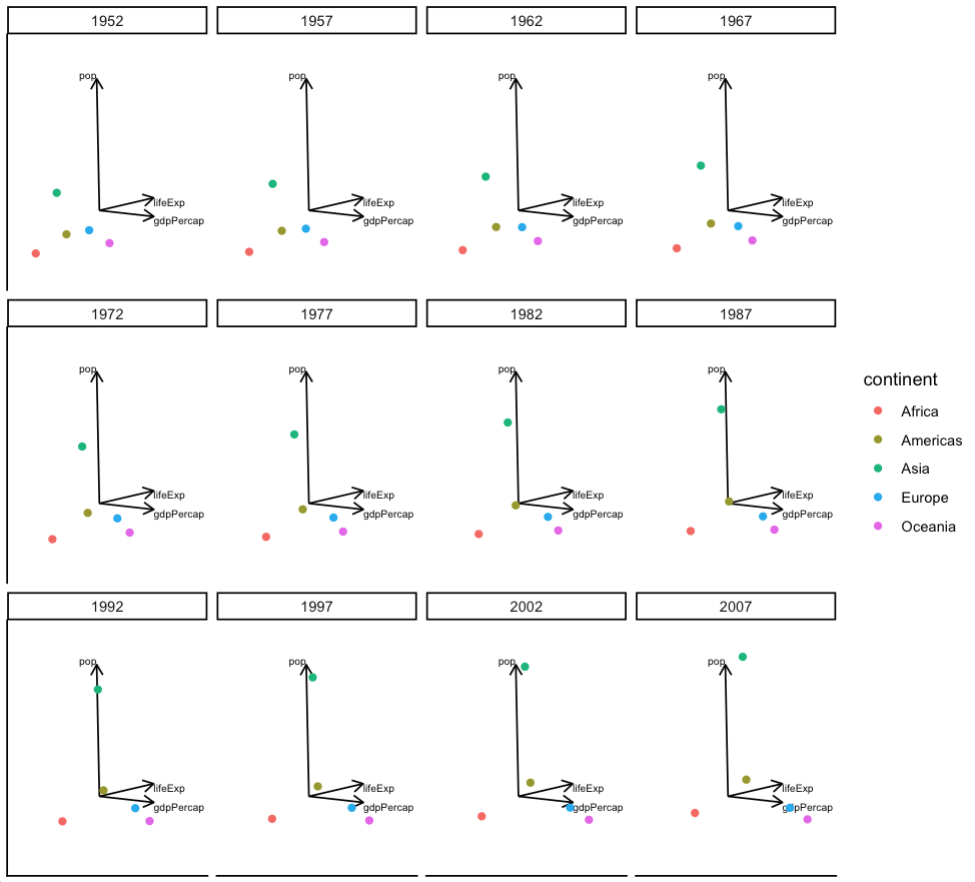} \caption{An animation (or facet) of the PCA biplot of the gapminder data using the function moveplot() under the fixed variable framework.}\label{fig:anim-plot}
\end{figure}

The animation (or facet) is shown in Figure \ref{fig:anim-plot} where the variable vectors remain fixed at their global positions while the continent points evolve across time, allowing the viewer to track how each continent moves relative to the directions of life expectancy, population, and GDP per capita. For instance, Asia and Africa show clear movement in the direction of increasing life expectancy and GDP over the decades, while population contributes less discriminatory information in the PCA space due to its large variance relative to the other variables. This is confirmed by the vector orientation of the variables.

It is worth noting that this is a deliberately simple example chosen to illustrate the mechanics of \texttt{moveplot()}. The broad trends visible in the animation, such as improvements in life expectancy and GDP for most continents, are also discernible in the static biplot of Figure \ref{fig:biplot-gap}, since the principal components are computed globally. The animation becomes more valuable in complex or less intuitive datasets, where temporal progression is difficult to perceive from a single static plot.

\subsection{\texorpdfstring{Demonstration: \emph{Africa climate} data}{Demonstration: Africa climate data}}\label{demonstration-africa-climate-data}

To illustrate \texttt{moveplot()} with multiple observations per group per time level, we use the \texttt{Africa\_climate} dataset included in \CRANpkg{moveEZ}. This dataset contains climate measurements for ten African regions derived from the ERA5 reanalysis (\citet{ERA5}), with IPCC-defined reference regions (\citet{IPCC}) used as the grouping variable. The dataset contains six continuous variables, described in Table \ref{tab:climate-table} alongside the abbreviations used in all biplot displays.

\begin{table}[!h]
\centering
\caption{\label{tab:climate-table}Variables in the Africa climate dataset and their biplot abbreviations.}
\centering
\resizebox{\ifdim\width>\linewidth\linewidth\else\width\fi}{!}{
\fontsize{9}{11}\selectfont
\begin{tabular}[t]{>{\raggedright\arraybackslash}p{3,5cm}>{\raggedright\arraybackslash}p{1.5cm}>{\raggedright\arraybackslash}p{2cm}>{\raggedright\arraybackslash}p{6cm}}
\toprule
Variable & Abbreviation & Unit & Description\\
\midrule
Accumulated Precipitation & AP & m/day & Total daily precipitation accumulated over the region\\
Daily Evaporation & DE & m/day & Net daily evaporation from the surface\\
Temperature & Temp & °C & Mean daily surface temperature\\
Soil Moisture & SM & m³/m³ & Volumetric water content of the upper soil layer\\
Standardised Precipitation Index (6-month) & SPI6 & Dimensionless & Standardised index of precipitation anomaly over a 6-month window\\
\addlinespace
Wind Speed & Wind & m/s & Mean daily wind speed at 10m above surface\\
\bottomrule
\end{tabular}}
\end{table}

Unlike the \emph{gapminder} example, each combination of region and time level contains many observations, allowing convex hulls to summarize regional spread at each time point.

\begin{figure}[!]
\includegraphics[width=1\linewidth]{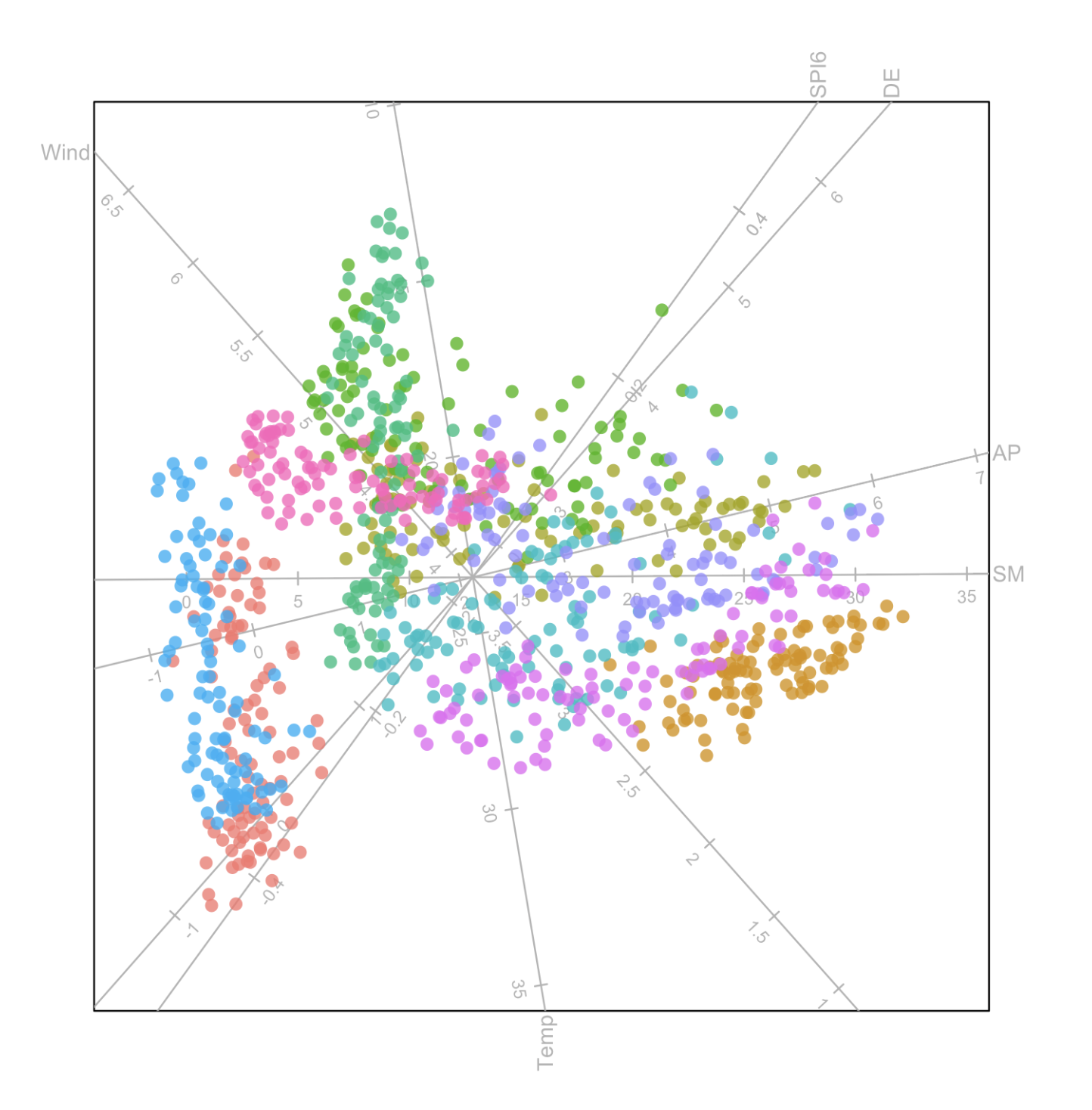} \caption{PCA biplot of the  climate data colored by region.}\label{fig:biplot-Afr}
\end{figure}

A PCA biplot is first constructed on the full dataset shown in Figure \ref{fig:biplot-Afr}.

\begin{verbatim}
bp <- biplot(Africa_climate, scaled = TRUE) |> 
  PCA(group.aes = Africa_climate$Region) |> 
  samples(opacity = 0.8, col = scales::hue_pal()(10)) |>
  plot()
\end{verbatim}

From the biplot in Figure \ref{fig:biplot-Afr}, the relationships between the variables are exposed, but there is some difficulty to separate the regions without intervention. Importantly, information on the time differences are not available which means a sequential exploration is impossible. These aspects are are addressed in \CRANpkg{moveEZ}.

Setting \texttt{move\ =\ FALSE} produces a faceted display, shown in Figure \ref{fig:moveplot-plot}, where convex hulls summarize each region at each time point:

\begin{verbatim}
bp |> moveplot(time.var = "Year", group.var = "Region", 
               hulls = TRUE, move = FALSE)
\end{verbatim}

\begin{figure}[!]
\includegraphics[width=1\linewidth]{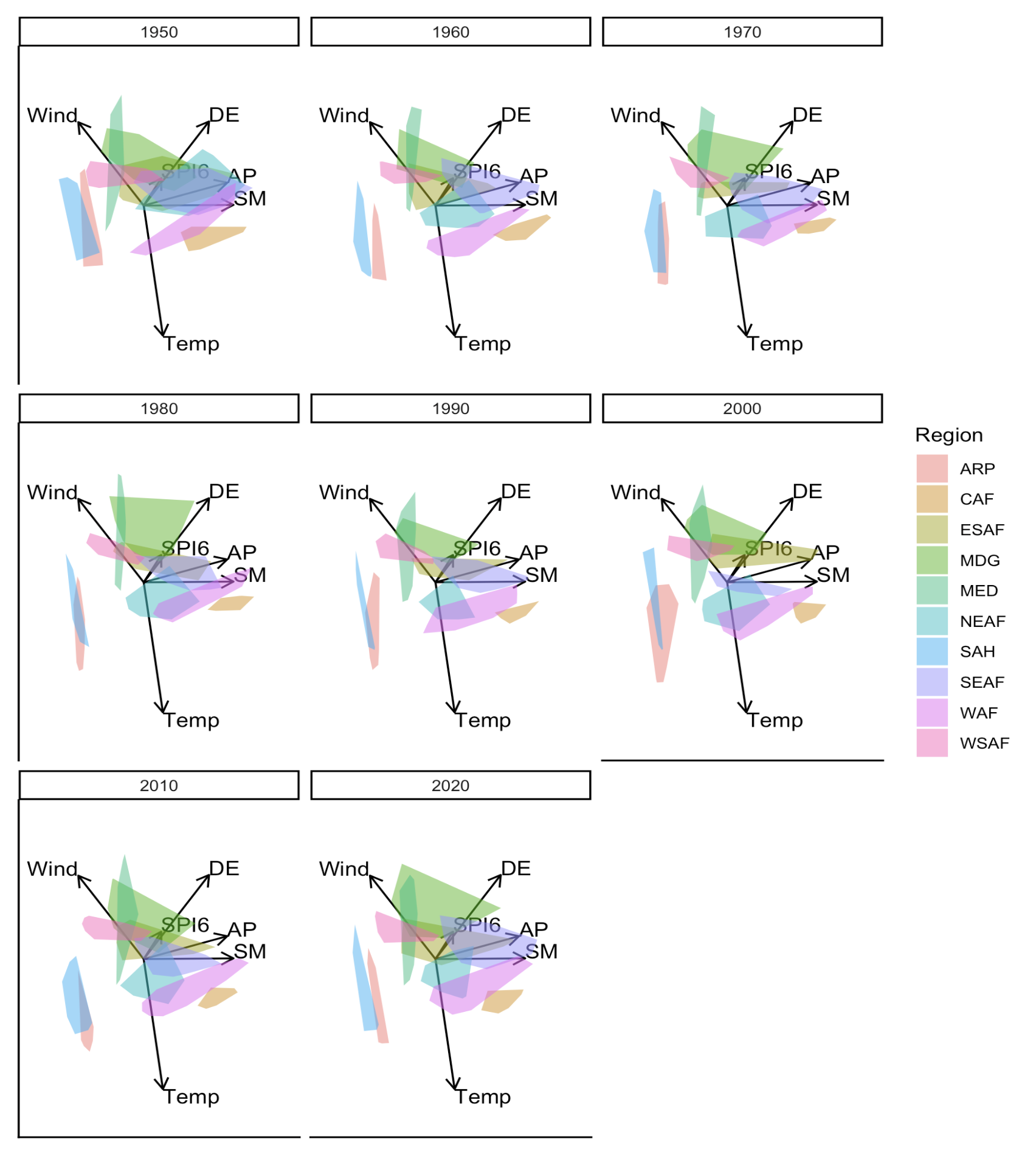} \caption{Faceted PCA biplots of the climate data, with each panel representing a distinct time frame, generated using moveplot().}\label{fig:moveplot-plot}
\end{figure}

Setting \texttt{move\ =\ TRUE} produces the animated equivalent, where the convex hulls evolve continuously across time, highlighting regional climate dynamics in a single integrated display:

\begin{verbatim}
bp |> moveplot(time.var = "Year", group.var = "Region", 
               hulls = TRUE, move = TRUE)
\end{verbatim}

Note that when fewer than three observations exist for a particular group and time level, sample points are plotted automatically in place of convex hulls, even when \texttt{hulls\ =\ TRUE} is specified.

\section{Dynamic frame}\label{dynamic-frame}

\begin{figure}[!]
\includegraphics[width=1\linewidth]{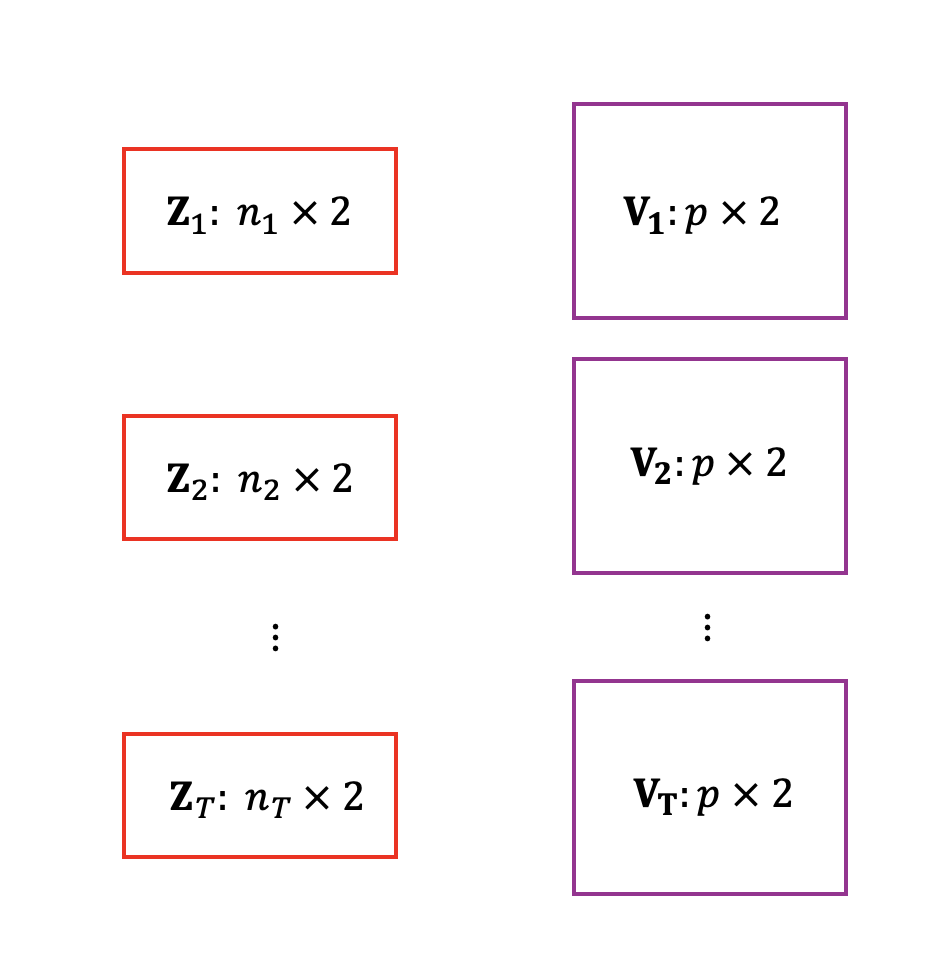} \caption{The dynamic variable framework where separate matrices ${\bf{Z}}$ and ${\bf{V}}$ are computed for each level of the time variable $T$.}\label{fig:dyn-frame}
\end{figure}

\texttt{moveplot2()} extends the animation to both sample coordinates and variable vectors by computing a separate PCA solution for each time slice of the data. Unlike \texttt{moveplot()}, which projects all time slices onto a fixed coordinate system defined by the full dataset, \texttt{moveplot2()} produces a separate pair of sample scores \({\bf{Z}}_t \, ; {\bf{V}}_t\) for \(t = 1, 2, \ldots, T\) as shown in Figure \ref{fig:dyn-frame}. Both sets of coordinates are then interpolated across time to form a continuous animation. This approach provides a more faithful depiction of time-varying variance--covariance structures and is particularly appropriate when the directional relationships among variables cannot be assumed stable across time.

Because \texttt{moveplot2()} fits a separate PCA per time slice, it requires a sufficient number of observations within each slice to perform the decomposition. It is therefore not suitable for datasets with a single observation per group per time level, such as the aggregated \emph{gapminder} data used in the previous section. The \texttt{Africa\_climate} dataset, which contains multiple observations per region and time level, is used to demonstrate this function throughout this section.

\begin{verbatim}
bp |> moveplot2(time.var = "Year", group.var = "Region", 
               hulls = TRUE, move = FALSE)
\end{verbatim}

\begin{figure}[!]
\includegraphics[width=1\linewidth]{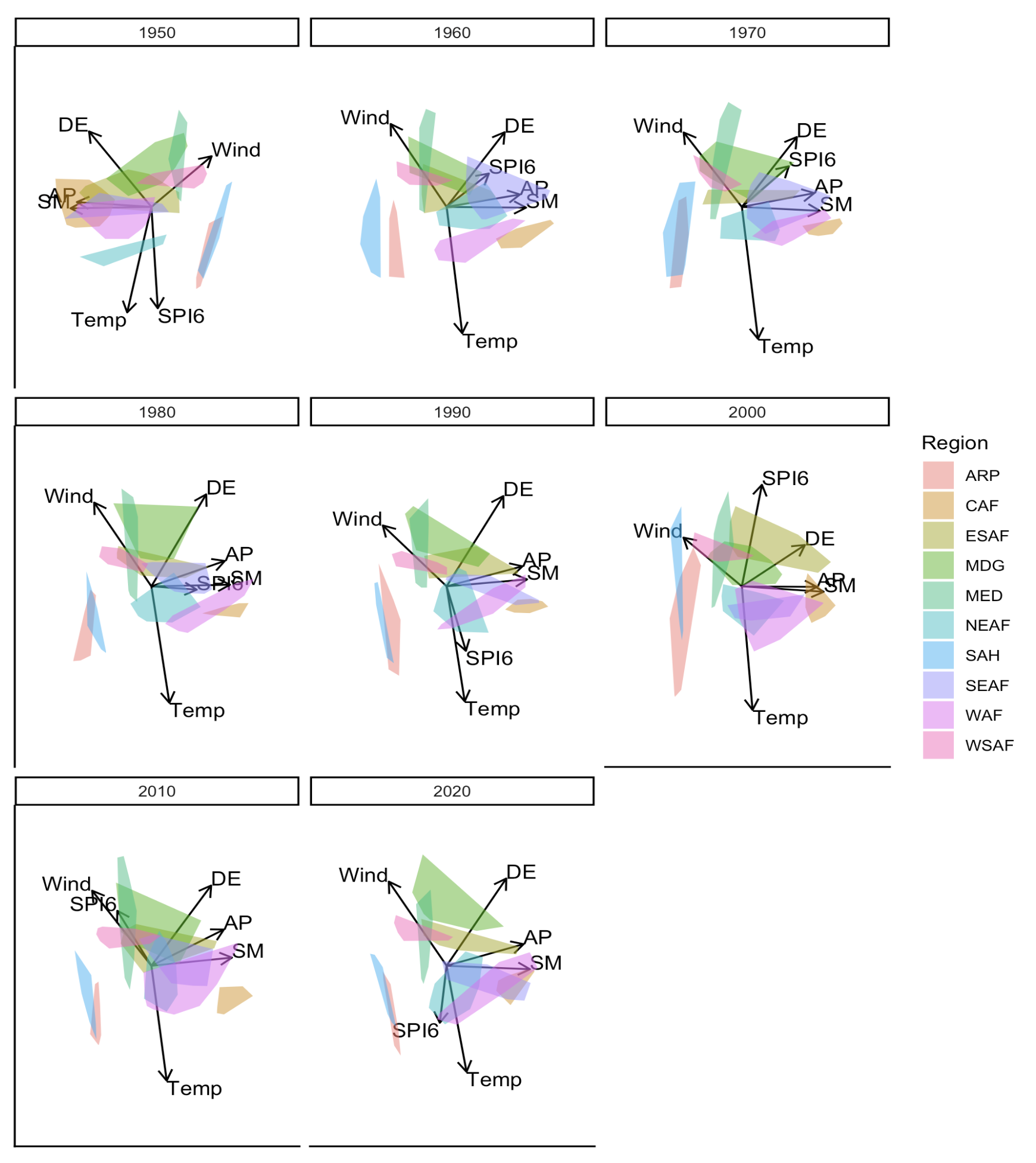} \caption{Faceted PCA biplots of the climate data using moveplot2().}\label{fig:moveplot-plot2}
\end{figure}

\subsection{Sign indeterminacy and visual discontinuity}\label{sign-indeterminacy-and-visual-discontinuity}

A consequence of computing separate PCA solutions per time slice is that the orientation of the resulting biplots is not uniquely defined. Eigenvectors are determined only up to a sign change, meaning that the same underlying covariance structure can be represented by configurations that are reflections of one another. This is mathematically inconsequential but visually disruptive: when the sign of an eigenvector flips between consecutive time slices, the variable vectors and sample positions may appear to suddenly mirror or reverse direction in the animation, creating a discontinuity that does not reflect any genuine change in the data.

This behavior can be seen in Figure \ref{fig:moveplot-plot2}, where a noticeable shift in orientation occurs between the biplots for 1950 and 1960, the variable vectors and sample configuration are reflected about the x-axis producing a visual jump in the animation. To correct for this, \texttt{moveplot2()} provides two optional arguments:

\begin{itemize}
\tightlist
\item
  \texttt{align.time}: a vector of time points at which alignment should be applied, ensuring that the orientation of principal components remains consistent across the specified slices.
\item
  \texttt{reflect}: detects and corrects sign inversions of the loadings between frames. Available options are \texttt{"x"} (reflect about the x-axis), \texttt{"y"} (reflect about the y-axis), and \texttt{"xy"} (reflect about both axes).
\end{itemize}

Applying these corrections to the 1950 slice will produce the aligned output.

\begin{verbatim}
bp |> moveplot2(time.var = "Year", group.var = "Region", 
               hulls = TRUE, move = FALSE,
               align.time = c("1950"),
               reflect = c("x"))
\end{verbatim}

While effective, this approach requires manual inspection to identify which time points need correction and how they should be reflected. For datasets with many time levels, this quickly becomes impractical, motivating the automated alignment strategy implemented in \texttt{moveplot3()}, described in the following section.

\section{Automated alignment}\label{automated-alignment}

The manual alignment approach provided by \texttt{moveplot2()} works well when the number of time levels is small and orientation discontinuities can be identified by visual inspection. However, as the number of time slices grows, determining the correct combination of \texttt{align.time} and \texttt{reflect} arguments becomes increasingly difficult and subjective. \texttt{moveplot3()} addresses this by automating the alignment of sequential biplots using GPA.

\subsection{Generalized Procrustes analysis}\label{generalized-procrustes-analysis}

Through GPA, multiple biplots are aligned by iteratively applying translation, reflection, rotation and scaling until the squared sum of squares distances between a set target biplot and the aligned biplots are minimized (\citet{Proc}). These transformations are mathematically inconsequential (i.e.~admissible) and preserves the distances between the coordinates. The \CRANpkg{GPAbin} package is used to perform GPA (\citet{GPAbinpack}). The target biplot typically represents the average of the coordinates of the biplots over the various time levels, which is represented by the additional argument \texttt{target\ =\ NULL}. However, it might be sensible for the specific application to compare the biplots to an additional time point which contains the same measurements, but for a different year. The example dataset in \CRANpkg{moveEZ} for this purpose, is \texttt{Africa\_climate\_target}, which represents measurements on the same variables as \texttt{Africa\_climate}, but for the year 1989.

\subsection{Demonstration: user-supplied target}\label{demonstration-user-supplied-target}

To compare each time slice in \texttt{Africa\_climate} against the 1989 reference measurements, the target dataset \texttt{Africa\_climate\_target} is first prepared:

\begin{verbatim}
viz_1989 <- Africa_climate_target |> 
  mutate("Target" = as.factor(rep("1989", nrow(Africa_climate_target))), 
         "Region" = as.factor(Region))

bp_1989 <- biplot(viz_1989, scaled = TRUE) |> 
  PCA(group.aes = viz_1989$Region)
\end{verbatim}

\begin{verbatim}
bp_1989 |> moveplot(time.var = "Target", 
  group.var = "Region", hulls = TRUE, move = FALSE)
\end{verbatim}

The resulting target biplot is used to align each time slice in \texttt{Africa\_climate} using GPA:

\begin{verbatim}
bp |> moveplot3(time.var = "Year", group.var = "Region", 
               hulls = TRUE, move = FALSE,
               target = Africa_climate_target)
\end{verbatim}

The faceted output will show each year's biplot aligned to the 1989 reference, exposing the structural differences between 1989 and each decade from 1950 to 2020.

\subsection{Demonstration: consensus target}\label{demonstration-consensus-target}

\begin{figure}[!]
\includegraphics[width=1\linewidth]{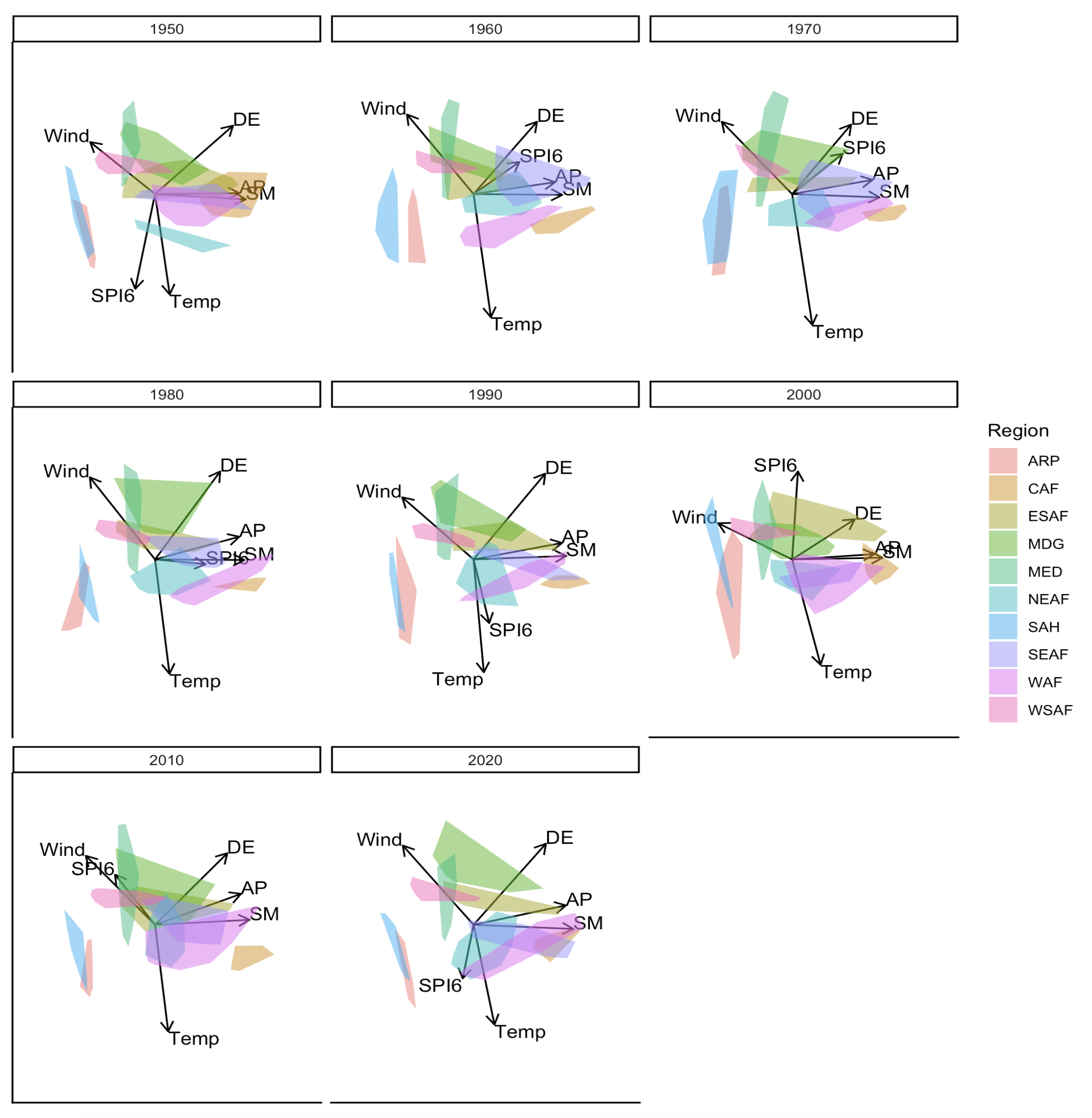} \caption{Faceted PCA biplots of the climate data using `moveplot3()` where no target is specified.}\label{fig:moveplot3-facet-plot}
\end{figure}

When no external reference is available or appropriate, setting \texttt{target\ =\ NULL} aligns all time slices to their average configuration as shown in Figure \ref{fig:moveplot3-facet-plot}.

\begin{verbatim}
bp |> moveplot3(time.var = "Year", group.var = "Region", 
               hulls = TRUE, move = FALSE, 
               target = NULL)
\end{verbatim}

Compared to the unaligned output of \texttt{moveplot2()} in Figure \ref{fig:moveplot-plot2}, the biplots are now consistently oriented across time, making it substantially easier to track genuine structural change rather than artifactual reorientation.

The next section will address the \texttt{evaluation} function in the package that assists in understanding the magnitude of the changes between biplots that are observed when applying \texttt{moveplot3()}.

\section{Evaluation}\label{evaluation}

The animated biplots produced by \texttt{moveplot3()} reveal how the multivariate structure of a dataset shifts relative to a target configuration across time. While these visual changes are informative, it is useful to quantify their magnitude, both to support objective interpretation and to identify time points that deviate most strongly from the target. The evaluation function in \CRANpkg{moveEZ} provides this by computing five measures of comparison between each individual biplot and the target configuration specified in \texttt{moveplot3()}, based on orthogonal Procrustes analysis between the two configurations. For a detailed treatment of these measures, refer to \citet{GPAbinart}.

\subsection{Measures of comparison}\label{measures-of-comparison}

The five measures fall into two categories. The fit measures assess the overall similarity between configurations:

\begin{itemize}
\tightlist
\item
  Procrustes Statistic (PS): values close to zero indicate high similarity between the two configurations.
\item
  Congruence Coefficient (CC): values close to one indicate high similarity.
\end{itemize}

The bias measures quantify the directional and magnitude discrepancy between configurations:

\begin{itemize}
\tightlist
\item
  Absolute Mean Bias (AMB): the average absolute displacement between matched coordinates.
\item
  Mean Bias (MB): the average signed displacement; a value close to zero indicates no systematic directional shift.
\item
  Root Mean Squared Bias (RMSB): a summary of overall coordinate-level discrepancy, penalizing large individual displacements more heavily than AMB.
\end{itemize}

Similarity between the target biplot and each biplot is reflected by small PS values (close to zero), large CC values (close to one) and low bias values (AMB, MB, RMSB) relative to others. These measures express the magnitude of changes that has to be made for a particular biplot to match the target visualization. Therefore, they measure how close the coordinates of the two configurations are.

\subsection{Extracting the measures}\label{extracting-the-measures}

The evaluation results are stored in the object returned by \texttt{moveplot3()} and can be extracted as follows:

\begin{verbatim}
results <- bp |> 
  moveplot3(time.var = "Year", group.var = "Region", hulls = TRUE, 
            move = FALSE, target = NULL) |> 
  evaluation()
results$eval.list
\end{verbatim}

Table \ref{tab:comp-tab} shows the measures for the first two comparisons, the target (1989) against 1950 and 1960, respectively. The higher PS and lower CC value for 1950 relative to 1960 indicates that the multivariate structure of 1950 is more dissimilar to the 1989 reference than that of 1960, which is consistent with the greater temporal distance between these years.

\begin{table}
\caption{\label{tab:comp-tab}Measures of comparison.}

\centering
\begin{tabular}[t]{lr}
\toprule
  & Target vs. 1950\\
\midrule
PS & 0.1323\\
CC & 0.9697\\
AMB & 1.2717\\
MB & 0.0000\\
RMSB & 1.8506\\
\bottomrule
\end{tabular}
\centering
\begin{tabular}[t]{lr}
\toprule
  & Target vs. 1960\\
\midrule
PS & 0.0982\\
CC & 0.9763\\
AMB & 0.4414\\
MB & 0.0000\\
RMSB & 0.5779\\
\bottomrule
\end{tabular}
\end{table}

\subsection{Visualizing the measures over time}\label{visualizing-the-measures-over-time}

When the number of time levels is large, inspecting a table of measures becomes unwieldy. \CRANpkg{moveEZ} therefore provides separate line plots for the fit and bias measures, which make temporal trends and anomalies immediately visible:

\begin{verbatim}
results$fit.plot
results$bias.plot
\end{verbatim}

\begin{figure}[!]
\includegraphics[width=1\linewidth]{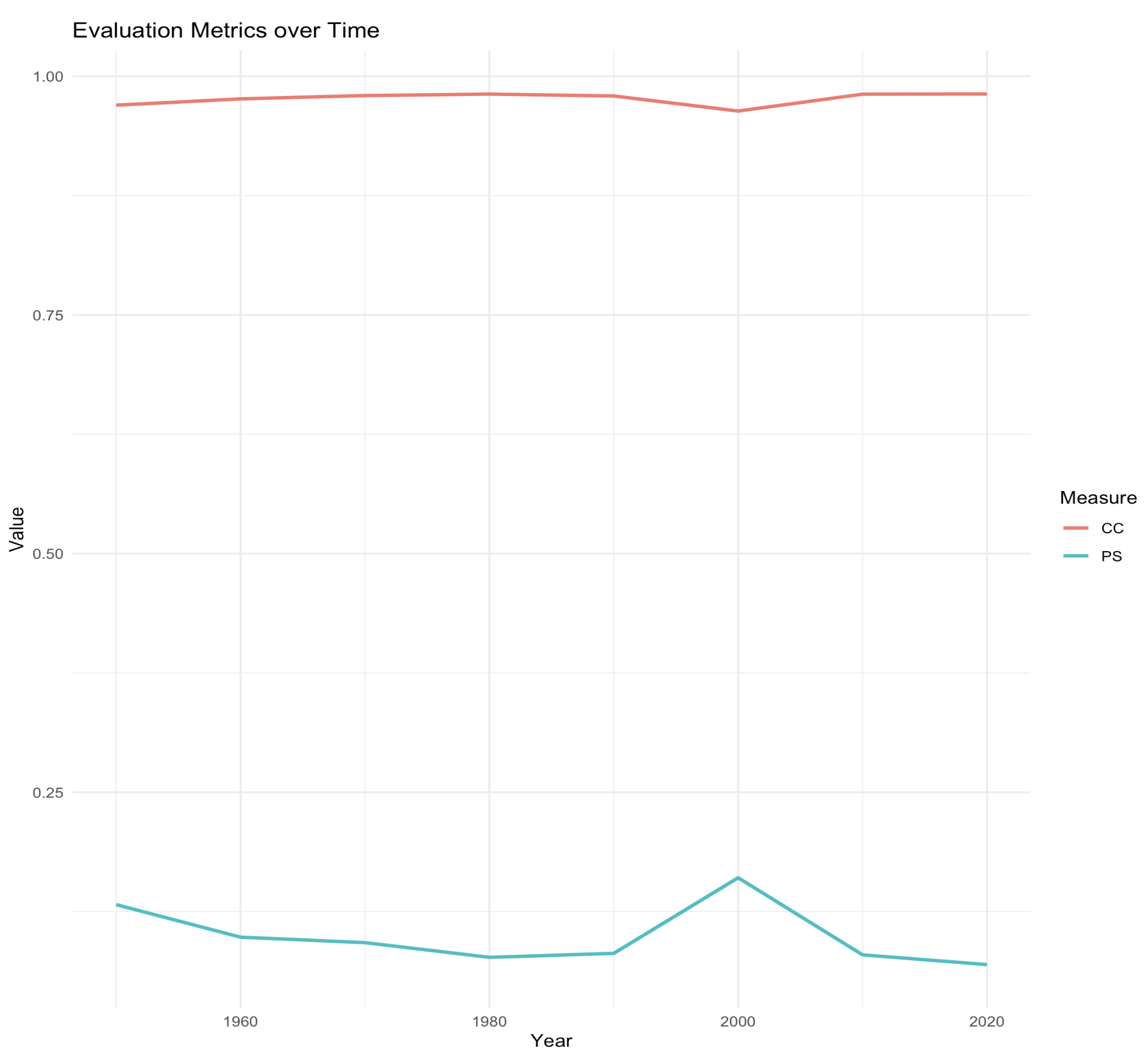} \caption{Fit measures using Procrustes Statistic (PS) and Congruence Coefficient (CC) for each time slice.}\label{fig:fit-line-plot}
\end{figure}

\begin{figure}[!]
\includegraphics[width=1\linewidth]{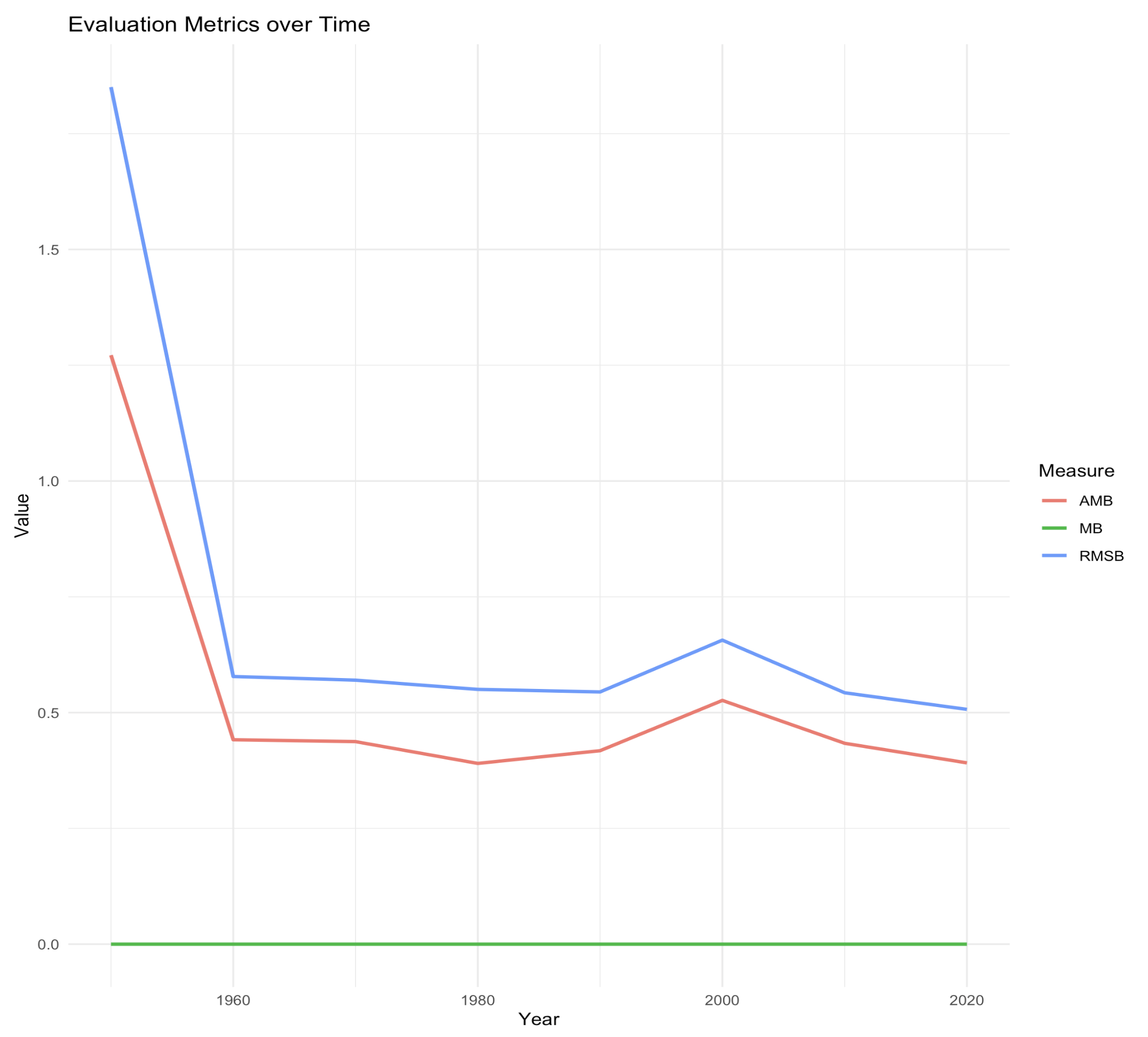} \caption{Bias measures using Absolute Mean Bias (AMB), Mean Bias (MB) and the Root Mean Squared Bias (RMSB) for each time slice.}\label{fig:bias-line-plot}
\end{figure}

The fit measures in Figure \ref{fig:fit-line-plot} reveal that the biplot of 2000 results in a lower CC and larger PS value compared to the other years. This means that there is a noticeable difference between the year 2000 and the average across years and the measurements of 2000 should be investigated in more detail to understand the cause of this difference.

The bias measures in Figure \ref{fig:bias-line-plot} shows that the initial bias is high, but decreases and stabilizes from 1960 with an increase in both the AMB and RMSB occurring for 2000. This is in agreement with the fit measures. The MB stays constant and close to zero for all comparisons.

\section{Conclusion}\label{conclusion}

The \CRANpkg{moveEZ} package addresses a gap in the existing R ecosystem by extending PCA biplot methodology to longitudinal and sequentially structured multivariate data. Prior to \CRANpkg{moveEZ}, analysts working with such data were limited to constructing separate static biplots per time point which is an approach that fragments temporal information and makes gradual structural change difficult to perceive. By animating transitions between time points and providing tools to quantify the magnitude of those changes, \CRANpkg{moveEZ} enables a more continuous and interpretable view of how multivariate structure evolves over time.

The three moveplot functions offer a coherent progression of methodological complexity. \texttt{moveplot()} provides an accessible entry point for datasets where the variance--covariance structure is stable across time or where only a single observation per group per time level is available. \texttt{moveplot2()} relaxes the assumption of a fixed coordinate system by computing a separate PCA solution for each time slice, at the cost of requiring manual correction of orientation discontinuities. \texttt{moveplot3()} automates this correction using GPA, making it the most robust option for datasets with many time levels or complex orientation behavior. The evaluation function complements all three by providing quantitative measures of structural similarity between time slices and a target configuration, supporting objective interpretation alongside the visual output.

Future development of \CRANpkg{moveEZ} will focus on extending animation support to the full range of biplot types available in \CRANpkg{biplotEZ}, beginning with canonical variate analysis and multiple correspondence analysis biplots, which share the most structural similarity with PCA biplots and are widely used in the applied sciences. A second avenue is the incorporation of non-linear easing functions, which would allow the speed of transitions between time states to reflect the magnitude of structural change rather than assuming a constant rate. Finally, the development of interactive controls such as a time slider, the ability to isolate individual group trajectories, or on-demand display of evaluation metrics alongside the animation would substantially enhance the utility of \CRANpkg{moveEZ} for exploratory data analysis in applied settings.

\CRANpkg{moveEZ} is available on CRAN and its development version is maintained on GitHub. Supplementary examples, including additional demonstrations of the animation frameworks and aesthetic customization, are provided in the accompanying package \href{https://cran.r-project.org/web/packages/moveEZ/vignettes/moveEZ.html}{vignette}.

\section{Acknowledgements}\label{acknowledgements}

We are grateful for the insights and contributions of Dianne Cook during the initial design phase of this package. Thank you for the financial support from the National Graduate Academy for Mathematical and Statistical Sciences (NGA(MaSS)).

\section{References}\label{references}

\bibliography{moveEZ.bib}

\address{%
Raeesa Ganey\\
University of Witwatersrand and Centre for Multi-Dimensional Data Visualisation (MuViSU)\\%
Johannesburg, South Africa\\
\textit{ORCiD: \href{https://orcid.org/0009-0008-6973-0999}{0009-0008-6973-0999}}\\%
\href{mailto:raeesa.ganey@wits.ac.za}{\nolinkurl{raeesa.ganey@wits.ac.za}}%
}

\address{%
Johané Nienkemper-Swanepoel\\
Stellenbosch University and Centre for Multi-Dimensional Data Visualisation (MuViSU)\\%
Stellenbosch, South Africa\\
\textit{ORCiD: \href{https://orcid.org/0000-0001-6086-8272}{0000-0001-6086-8272}}\\%
\href{mailto:nienkemperj@sun.ac.za}{\nolinkurl{nienkemperj@sun.ac.za}}%
}